\documentclass[conference, 10pt]{IEEEtran}
\IEEEoverridecommandlockouts
\usepackage{cite}
\usepackage{amsmath,amssymb,amsfonts}
\usepackage{algorithmic}
\usepackage{graphicx}
\usepackage{textcomp}
\usepackage{bbm}
\usepackage{xcolor}
\def\BibTeX{{\rm B\kern-.05em{\sc i\kern-.025em b}\kern-.08em
    T\kern-.1667em\lower.7ex\hbox{E}\kern-.125emX}}

\usepackage{tikz}
\usepackage{amsmath,amssymb,color,graphicx,caption,subcaption,comment,tikz,url,latexsym} 
\usepackage{algorithmic}
\usepackage{graphicx}
\usepackage{textcomp}
\usepackage{pgfplots}
\usepackage{xcolor}
\usepackage{url}
\usepackage{multirow}
\usepgfplotslibrary{units}
\usetikzlibrary{spy,backgrounds}
\usepackage{pgfplotstable}

\usepackage[T1]{fontenc}

\usepackage{dsfont} 

\usepackage{bm} 

\newtheorem{theorem}{Theorem}

\allowdisplaybreaks[4]
\newcommand{\highlight}[1]{{\color{black}#1}}
\newcommand{\ab}[1]{{\color{black}#1}}

\newcommand{\mw}[1]{{\color{black}#1}}

\begin{document}

\title{Covert Multi-Access Communication with a Non-Covert User}

    \author{
    \IEEEauthorblockN{Abdelaziz Bounhar$^{\star}$, Mireille Sarkiss$^{\S}$, Mich\`ele Wigger$^{\star}$}
        \IEEEauthorblockA{$^{\star}$LTCI, T\'{e}l\'{e}com Paris, Institut Polytechnique de Paris, 91120 Palaiseau, France
    \\\{abdelaziz.bounhar, michele.wigger\}@telecom-paris.fr}
        \IEEEauthorblockA{$^{\S}$SAMOVAR, T\'{e}l\'{e}com SudParis, Institut Polytechnique de Paris, 91120 Palaiseau, France
	\\\{mireille.sarkiss\}@telecom-sudparis.eu}
}

\maketitle

\begin{abstract}
In this paper, we characterize the fundamental limits of a communication system with three users (i.e., three transmitters)  and a single receiver where  communication from two covert users  must remain undetectable to an external warden. Our results show a tradeoff between the highest rates that are simultaneously achievable for the three users. 
They further show that  the presence of a non-covert user  in the system can enhance the capacities of the covert users under stringent secret-key constraints.
To derive our fundamental limits, we provide  an information-theoretic converse  proof and present a coding scheme that achieves the performance of our converse result.  Our coding scheme is based on 
 multiplexing different code phases, which seems to be  essential to exhaust the entire tradeoff region between the rates at the covert and the two non-covert users. This property is reminiscent of the setup with multiple non-covert users, where multiplexing is also required to exhaust the entire rate-region. 
\end{abstract}


\section{Introduction}
\label{section:introduction}
\IEEEPARstart{G}{}uaranteeing privacy and security of Internet of Things (IoT) and sensor networks is a major challenge for future wireless systems~\cite{iot_security_survey}. In many IoT applications, devices are resource-constrained and transmit sporadically a small number of bits while remaining silent for most of the time. 
\highlight{Such transmissions can be secured through the paradigm of} covert communication, a physical layer security technique, where users convey information without being detected by external wardens.
\highlight{It} was shown in \cite{bash_first} that it is possible to communicate covertly as long as the number of  communicated bits scales like $\mathcal{O}(\sqrt{n})$, for $n$ the number of channel uses, which is compliant with IoT scenarios. 
Covert-rates were first characterized according to \highlight{this} so-called \emph{square-root law} over AWGN channels in \cite{bash_first}. 
Several subsequent works~\cite{bash_p2p, bloch_first, ligong_first}, made this \emph{square-root law} become the de-facto standard limit of covert communication for most scenarios and channels.
Extensions to Broadcast Channels (BCs) and Multiple Access Channels (MACs) were proposed in  \cite{bloch_journal_embedding_broadcast, ligong_broadcast, sang_multiple_overt_superposition_covert_on_overt,  ours_first}.

This paper generalizes our previous work \cite{ours_first} to a Discrete Memoryless Multiple Access Channel (DM-MAC) with two covert users communicating with a legitimate receiver without being detected by an external warden, while a third non-covert user is not subject to \highlight{any} such covertness constraint.
\highlight{We establish the fundamental limits  of all achievable tuples of non-covert rate, covert rates, and secret-key rates.}

Our results show a fundamental tradeoff between the rates \highlight achievable by all users. They also confirm our previous conclusions in~\cite{ours_first} on the fundamental role of  multiplexing different code strategies to exhaust the fundamental tradeoff between the covert and non-covert rates, and on the benefits of the non-covert user to improve the covert users' capacity \highlight{under a secret-key rate constraint}.

\section{Notation}
We follow standard notations in \cite{ours_first, cover, Csiszarbook}. In particular, we denote a random variable by $X$ and its realization by $x$. 
We write $X^{n}$ and $x^{n}$ for the tuples $(X_1,\ldots, X_n)$ and $(x_1,\ldots, x_n)$, respectively, for any positive integer $n > 0$. 
For a distribution $P$ on $\mathcal{X}$, we note its product distribution on $\mathcal{X}^{n}$ by $P^{\otimes n}(x^{n}) = \prod_{i=1}^{n} P(x_{i})$. 
For two distributions $P$ and $Q$ on $\mathcal{X}$, $\mathbb{D}(P\|Q)=\sum_{x \in \mathcal{X}} P(x)\log(\frac{P(x)}{Q(x)})$ denotes the Kullback-Leibler divergence between $P$ and $Q$.

\section{Problem statement}
\label{sec:problem_statement}
\begin{figure}[t!]
   \centering
   \includegraphics[scale=0.75]{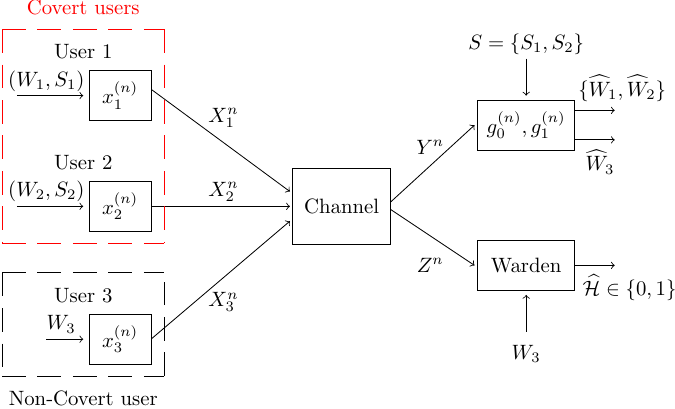}
    \caption{MAC setup with 2 covert users and a non-covert user in the presence of  an external warden.} 
   \label{fig:setup}
\end{figure}
\label{section:problem_setup_and_main_result}
Consider the three-user single-receiver setup in Figure~\ref{fig:setup} where an external warden should not be able to detect communication from 
 Users 1 and 2. Communication from User 3 has no detectability constraints, and we can even allow the warden to know its transmitted message.\footnote{Providing the warden with the  message of User 3 makes the warden only stronger. The rates that are achievable under such a strong warden remain also achievable under weaker assumptions on the warden.} 
We model our setup using two hypotheses $\mathcal{H}=0$ and $\mathcal{H}=1$, where under $\mathcal{H}=0$ only User 3 is transmitting  while under $\mathcal{H}=1$ all three users transmit, and the warden wishes to guess the true hypothesis. Details are as follows. For simplicity of illustration, we assume that Users 1 and 2 produce inputs in the binary alphabets $\mathcal{X}_1=\mathcal{X}_2=\{0,1\}$ and we consider that $0$ is the "off-symbol", i.e. the symbol transmitted by Users 1 and 2 under $\mathcal{H}=0$. User~3's input alphabet $\mathcal{X}_3$ is finite but arbitrary otherwise. 
The legitimate receiver and the warden observe channel outputs in the finite alphabets $\mathcal{Y}$ and $\mathcal{Z}$. 
Define the message and key sets 
\begin{IEEEeqnarray}{rCl}
\mathcal{M}_{\ab{\ell}} &\triangleq& \{1, \ldots, \mathsf{M}_{\ab{\ell}} \}, \quad \forall \ab{\ell} \in \{1,2,3\},\\
\mathcal{K}_\ab{\ell} & \triangleq & \{1,\ldots, \mathsf{K}_\ab{\ell}\}, \quad \forall \ab{\ell} \in \{1,2\},
\end{IEEEeqnarray}
for given numbers $\mathsf{M}_1$,  $\mathsf{M}_2$, $\mathsf{M}_3$, $\mathsf{K}_1$, and $\mathsf{K}_2$ and let the messages $W_1$, $W_2$, $W_3$ and the keys $S_1$ and $S_2$ be independent of each other and uniform over $\mathcal{M}_1, \mathcal{M}_2$, $\mathcal{M}_{3}$, $\mathcal{K}_1$ and $\mathcal{K}_2$, respectively. 
For each $\ab{\ell} \in \{1,2\}$, the key $S_\ab{\ell}$ is known to User $\ab{\ell}$ and to the legitimate receiver, while message $W_\ab{\ell}$ is known to User $\ab{\ell}$ only. 
In contrast $W_3$ is known to User 3 and is given to the warden. \\
\textit{\underline{Under $\mathcal{H}=0$:}} Users 1 and 2 send the all-zero sequences 
\begin{align}
X_\ab{\ell}^n&=0^n, \quad \forall \ab{\ell} \in \{1,2\},
\end{align} 
whereas User 3 applies an encoding function $x_3^{(n)}\colon \mathcal{M}_3 \to \mathcal{X}_3^n$ to its message $W_3$ and sends the resulting codeword 
\begin{equation}\label{eq:X3}
X_3^n=x_3^{(n)}(W_3)
\end{equation} over the channel.\\
\textit{\underline{Under $\mathcal{H}=1$:}} For each $\ab{\ell} \in \{1,2\}$, User $\ab{\ell}$ applies an encoding function $x_\ab{\ell}^{(n)}\colon \mathcal{M}_\ab{\ell}\times \mathcal{K}_\ab{\ell}\to \mathcal{X}_\ab{\ell}^n$ to its message $W_\ab{\ell}$ and to the secret key $S_\ab{\ell}$ and sends the resulting codeword 
\begin{equation}
X_\ab{\ell}^n=x_\ab{\ell}^{(n)}(W_\ab{\ell},S_\ab{\ell}), \quad \forall \ab{\ell} \in \{1,2\},
\end{equation} over the channel. 
User 3, unaware of whether $\mathcal{H}=0$ or $\mathcal{H}=1$, constructs its channel inputs as in~\eqref{eq:X3}.

The legitimate receiver,  which knows the hypothesis $\mathcal{H}$, decodes the desired messages $W_3$ (under $\mathcal{H}=0$) or  $(W_1,W_2,W_3)$ (under $\mathcal{H}=1$) based on its knowledge of the \ab{secret-}keys $(S_1,S_2)$ and its observed outputs $Y^n=(Y_1,\ldots, Y_n)$ which are generated by a discrete memoryless channel \highlight{$\Gamma_{Y \mid X_1X_2X_3}$} from the input sequences $X_1^n, X_2^n, X_3^n$.
That means, if $X_1^n=x_1^n$, $X_2^n=x_2^n$, and $X_3^n=x_3^n$ then the $i$-th output symbol $Y_i$ is generated from the $i$-th inputs $x_{1,i}, x_{2,i}, x_{3,i}$ according to the conditional channel law \highlight{$\Gamma_{Y \mid X_1X_2X_3}(\cdot | x_{1,i}, x_{2,i}, x_{3,i})$} for any $i \in \{1, \ldots, n\}$.

\highlight{Under $\mathcal{H}=0$, the decoder} uses a decoding function $g_0^{(n)}\colon \mathcal{Y}^n \to \mathcal{M}_3$ to produce the single guess
\begin{equation}
\widehat{W}_3 = g_0^{(n)}( Y^n)
\end{equation}
and under $\mathcal{H}=1$ it uses a decoding function $g_1^{(n)}\colon \mathcal{Y}^n \times \mathcal{K}_1 \times \mathcal{K}_2 \to \mathcal{M}_1 \times  \mathcal{M}_2 \times  \mathcal{M}_3$ to produce the triple of guesses
\begin{equation}
(\widehat{W}_1,\widehat{W}_2, \widehat{W}_3) = g_1^{(n)}( Y^n, S_1, S_2). 
\end{equation}
The decoder performance associated with a tuple of encoding and decoding functions $(x_1^{(n)}, x_2^{(n)}, x_3^{(n)}, g_0^{(n)}, g_1^{(n)})$ is measured by the error probabilities under the two hypotheses:
\begin{IEEEeqnarray}{rCl}
    \ab{P_{e,0}} & \triangleq & \Pr\left(\widehat{W}_3 \neq W_3 \Big| \mathcal{H}=0\right), \label{eq:prob2} \\
    \ab{P_{e,1}} &\triangleq& \Pr\left(\bigcup_{\ab{\ell}=1}^3 \widehat{W}_\ab{\ell} \neq W_\ab{\ell} \Big| \mathcal{H}=1\right). \label{eq:prob1}
\end{IEEEeqnarray}

On the other side, the warden observes the message $W_3$ and the channel outputs $Z^n$, which are generated from inputs $X_1^n, X_2^n, X_3^n$ according to an arbitrary but given discrete and memoryless channel law $\Gamma_{Z \mid X_1X_2X_3}$.
By the uniform nature of the messages and the \ab{secret-}keys, for each $w_3\in \mathcal{M}_3$ and $W_3=w_3$, the warden's output distribution under $\mathcal{H}=1$ is  
\begin{multline}
    \label{eq:def_Q_C_w3}
        \widehat{Q}_{\mathcal{C}, w_3}^{n}(z^{n}) \triangleq  \frac{1}{
        \mathsf{M}_1\mathsf{M}_2\mathsf{K}_1\mathsf{K}_2} \Bigg[ \sum_{(w_1,s_1)} \sum_{(w_2,s_2)} \\
        \Gamma^{\otimes n}_{Z|X_1X_2X_3} (z^n| x_1^n(w_1,s_1), x_2^n( w_2,s_2), x_3^n(w_3))\Bigg],
\end{multline}
and under $\mathcal{H}=0$, it is
\begin{IEEEeqnarray}{rCl}
\Gamma^{\otimes n}_{Z|X_1X_2X_3} (z^n| 0^n, 0^n, x_3^n(w_3)).
\end{IEEEeqnarray}
For any $w_3 \in \mathcal{M}_3$, the covertness constraint at the warden is defined by means of  the divergence
\begin{equation}
\label{eq:def_delta_n_w3}
\delta_{n,w_3}\triangleq \mathbb{D}\left(\widehat{Q}_{\mathcal{C}, w_3}^{n} \big\| \Gamma^{\otimes n}_{Z|X_1X_2X_3} ( \cdot | 0^n, 0^n, x_3^n(w_3))\right).
\end{equation}
This divergence  can be related to the warden's detection error probabilities by standard arguments \cite[Section 11.8]{cover}.

\section{Coding Scheme, Main Result, and Numerical Simulations}
\label{sec:coding_and_simulation}

\subsection{Coding Scheme}
\label{sec:coding_overview}
Our coding scheme multiplexes $\tau$ different\footnote{We will see that \highlight{$\tau=6$ suffices.}} phases and codes  $t=1, 2, \ldots, \tau$. The need of multiple phases stems from  the multi-objective nature of our communication that not only wishes to minimize various error probabilities but also the divergence between the output distribution observed at the warden under the two hypotheses (so as to reduce the warden's detection capability). While certain phases will provide small probabilities of error, others  will induce small divergences. The combined scheme over all phases  then induces an optimal overall-tradeoff between small probabilities of error and small divergences.

\textit{\underline{Code construction:}} Our code construction has the following parameters: 
\begin{itemize}
\item a sequence of positive numbers $\{\omega_n\}_{{n \in \mathbb{N}}}$ satisfying
\begin{subequations}   
    \label{eq:lim_omega_n_def}
    \begin{IEEEeqnarray}{rCl}
        \lim_{n \rightarrow \infty} \omega_{n} &=& 0, \\  
          \lim_{n \rightarrow \infty}\left( \omega_{n}\sqrt{n} -\log n\right) &=& \infty;
    \end{IEEEeqnarray}
\end{subequations}
\item  a probability distribution $P_T$ over $\{1,\ldots, \tau\}$;
\item non-negative values  $\{ \rho_{1,t}, \rho_{2,t}\}_{t=1}^\tau$;
\item {conditional} probability distributions {$P_{X_3 \mid T=t}$ over $\mathcal{X}_3$}, for $t=1, \ldots, \tau$. 
\end{itemize}

For any blocklength $n$ we split the entire blocklength $n$ into $\tau$ transmission phases $t=1,2,\ldots, \tau$, where the $t$-th phase is of length \ab{$n_t:= \lfloor n \cdot P_T(t) \rfloor $}. 

We pick pairs of  non-detectable multiple-access codes  \cite{bloch_k_users_mac} $\{ \mathcal{C}_{1,t}, \ \mathcal{C}_{2,t}\}$ for $t=1,\ldots, \tau$, for the transmission of the two covert messages $W_1$ and $W_2$. By construction \cite{bloch_k_users_mac, tin_paper, ours_first}, codebooks  $\mathcal{C}_{1,t}$ and $\mathcal{C}_{2,t}$ contain codewords $\big\{x_{1,t}^{n_t}(W_1, S_1)\big\}$ and $\big\{x_{2,t}^{n_t}(W_2, S_2)\big\}$ that depend on the respective messages as well as the corresponding \ab{secret-}keys $S_1$ and $S_2$.

An important parameter of the covert-communication codes $\mathcal{C}_{1,t}$ and $\mathcal{C}_{2,t}$ is the average number of $1$-symbols in the codewords. For each $t=1, \ldots, \tau$, we  choose each codebook $\mathcal{C}_{1,t}$  to consist of  codewords  containing approximately  $\rho_{1,t}\omega_n \sqrt{n}_t$  $1$-symbols and $\mathcal{C}_{2,t}$  to consist of  codewords  containing approximately  $\rho_{2,t}\omega_n \sqrt{n}_t$ $1$-symbols.

Standard (non-covert) single-user codes $\mathcal{C}_{3,t}$, for $t=1,\ldots, \tau$ are used  for the transmission of the non-covert message $W_3$ in the different phases. The codewords $x_{3,t}^{n_t}(W_3)$ depend only on message $W_3$. A key parameter of these codes is again the frequency of the various symbols in the codewords, which we call  $P_{X_3 \mid T=t}$ for codebook $\mathcal{C}_{3,t}$. It is  fixed and independent of the blocklength.

\textit{\underline{Encoding:}} User $3$ forms the concatenation of codewords 
\begin{equation}
x_3^{3}(W_3):= x_{3,1}^{n_1}(W_3) , \; x_{3,2}^{n_2}(W_3),\;  \ldots, \; x_{3,\tau}^{n_\tau}(W_3)
\end{equation}
and sends the resulting string over the channel.

Under $\mathcal{H}=0$, Users 1 and 2 send the all-zero sequences $x_1^n=0^n$ and $x_2^n=0^n$. Under $\mathcal{H}=1$, Users 1 and 2 concatenate the codewords from the different codebooks 
\begin{eqnarray}
\highlight{\lefteqn{x_{\ab{\ell}}^{n}(W_\ab{\ell},S_\ab{\ell}) : = }} \nonumber \\
& x_{\ab{\ell},1}^{n_1}(W_\ab{\ell},S_\ab{\ell}) , \; x_{\ab{\ell},2}^{n_2}(W_\ab{\ell},S_\ab{\ell}),\;  \ldots, \; x_{\ab{\ell},\tau}^{n_\tau}(W_\ab{\ell},S_\ab{\ell}), \nonumber \\
&\hspace{5cm} \qquad \forall  \ab{\ell}\in \{1,2\} \IEEEeqnarraynumspace
\end{eqnarray}
and send the resulting strings over the channel. 

\begin{figure}[t!]
   \centering
    \includegraphics[scale=0.7]{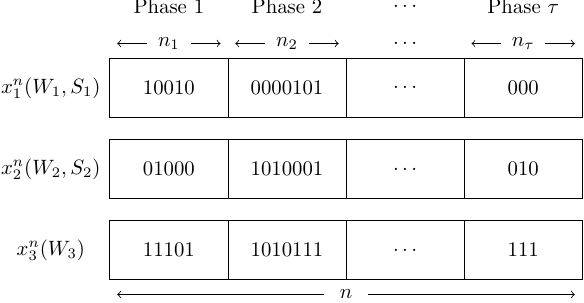}
    \caption{Our encoding process under $\mathcal{H}=1$ for binary input alphabets at all users. Under $\mathcal{H}=0$, the covert users 1 and 2 send the all-zero sequence.}
    \label{fig:encoding_process}
\end{figure}

The encoding process under $\mathcal{H}=1$ is depicted in Figure~\ref{fig:encoding_process}. Notice that the number of 1-symbols varies from one user to the other and it also varies over the $\tau$ phases.  In particular, the covertness constraint imposes on Users 1 and 2 to transmit a limited number of 1-symbols thereby making the codewords sparse, which is not the case for User 3.

\textit{\underline{Decoding:}}
The legitimate receiver employs the following successive decoding procedure: 
\begin{enumerate}
\item The receiver first decodes the non-covert message $W_3$ based on the entire output sequence $Y^n$ and using all $\tau$ codebooks $\mathcal{C}_{3,1}, \ldots, \mathcal{C}_{3,\tau}$, and assuming that Users 1 and 2 send the all-zero sequences.   Various decoding algorithms can be employed, for example joint typicality decoding or a maximum likelihood decoding rule based on \emph{all the $\tau$ codebooks}. 

During this decoding steps it is assumed that both  Users 1 and 2 send the all-zero codewords, irrespective of whether $\mathcal{H}=0$ or $\mathcal{H}=1$. In fact, even under  $\mathcal{H}=1$ this assumption will not distort the problem too much because the number of $1$-symbols is small (in the order of $\omega_n\sqrt{n}$) anyways.

\item Having decoded  the message $W_3$, the receiver proceeds with decoding either message $W_1$ or message  $W_2$. Assume it starts by decoding $W_1$ using a standard decoding rule based on all $\tau$ covert-codes $\{\mathcal{C}_{1,t}\}$. For this decoding step the receiver can  assume that $x_2^n=0^n$.  A potential decoding rule is to combine the likelihoods of the $\tau$ codewords to take a decision on the transmitted message $W_1$. A second alternative is to look for an index $j$ satisfying
\begin{equation}
\log \left(\frac{ \highlight{\Gamma_{Y|X_1X_2X_3}^{\otimes n}}  (Y^n | x_1^n(\highlight{j},S_1), \ab{0^n}, X_{3}^n\highlight{(\widehat{W}_3)}) }{\highlight{\Gamma^{\otimes n}_{Y|X_1X_2X_3}}(Y^n| \ab{0^n}, \ab{0^n}, X_{3}^n\highlight{(\widehat{W}_3)})} \right) > \eta_1,
\end{equation}
for a suitably chosen constant $\eta_1$. \highlight{Here, $\widehat{W}_3$ denotes the receiver's guess of $W_3$.}

If such an index exists and is  unique, the receiver declares $W_1$ to be equal to this index. Otherwise it declares and error and stops. 
\item The receiver proceeds to decode the second covert message $W_2$ in a way that is analogous to the decoding of message $W_1$, but now $x_1^n$ \highlight{(and not $x_2^n$)} is assumed to be the all-zero codeword. 
\end{enumerate}

\begin{figure}[t!]
   \centering
    \includegraphics[scale=0.8]{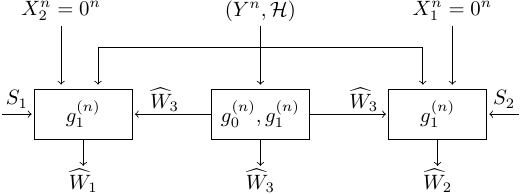}
    \caption{Under $\mathcal{H}=1$, the non-covert message $W_3$ is decoded first, followed by parallel decoding of the covert messages using an Interference as Noise scheme. Under $\mathcal{H}=0$, we decode only $W_3$.}
    \label{fig:decoding_process}
    \vspace{-3mm}
\end{figure}

Steps 2) and 3) can be inverted or even be ran in parallel. It is however important that Step 1) is executed first. The decoding process is also depicted in  Figure~\ref{fig:decoding_process}.
\mw{\subsection{Generalization of the Coding Scheme}\label{sec:gen}
We propose a slight generalization of our coding scheme including two new parameters $\phi_1 , \phi_2 \in (0,1]$. In our description, we assume $\phi_1 \geq \phi_2$, otherwise we will switch the roles of Users 1 and 2. 

In the generalized scheme, communication at Users 1 and 2 is only over a fraction $\phi_1$ of each phase; during the remaining $(1-\phi_1)$ fraction of each phase both users simply send the all-zero symbols. User 3 acts as before. Specifically, during the first $\phi_1$ fraction of each phase, User 1 communicates as described in the previous section, and accordingly, the codebooks $\{\mathcal{C}_{1,t}\}$ contain codewords of lengths $n_t \phi_1$. Instead, codebooks $\{\mathcal{C}_{2,t}\}$ contain codewords of length $n_t \phi_2< n_t \phi_1$. In fact, User 2 sends the appropriate codeword from these codebooks during the first $n_t \phi_2$ channel uses of each phase, and during the next $n_t (\phi_1-\phi_2)$ channel uses it sends i.i.d. symbols $\{X_{2,i}\}$ drawn according to the same distribution as used in the code construction of the phase.  This ensures a homogeneous expected divergence at the warden  under the two hypotheses during the first $n_t \phi_1$ channel uses of each phase. During the last $(1-\phi_1)$ fraction of each phase the divergence is zero  because under both hypotheses Users 1 and 2 both send the all-zero symbol. 

It can be shown that the described modifications yield a factor $\phi_1$ for the logarithmic message size that can be reliably transmitted at User 1 and for the divergence, while they yield a factor $\phi_2$ for the logarithmic message size that can be reliably sent by User 2. 
}
\subsection{Main Results}
For ease of notation, define \highlight{\begin{IEEEeqnarray}{rCl}
\Gamma_{x_1x_2x_3}^{Y}(y) &\triangleq& \Gamma_{Y|X_1X_2X_3}(y \mid x_1, x_2, x_3),\\
\Gamma_{x_1x_2x_3}^{Z}(z) &\triangleq& \Gamma_{Z|X_1X_2X_3}(z \mid x_1, x_2, x_3),
\end{IEEEeqnarray}}
and
\begin{IEEEeqnarray}{rCl}
D_{Y}^{(1)}(x_3)& \triangleq & \mathbb{D} \left(\Gamma_{10x_3}^{Y} \mid \mid \Gamma_{00x_3}^{Y}  \right), \\
D_{Y}^{(2)}(x_3)& \triangleq & \mathbb{D} \left(\Gamma_{01x_3}^{Y} \mid \mid \Gamma_{00x_3}^{Y}  \right), \\
D_{Z}^{(1)}(x_3)& \triangleq & \mathbb{D} \left(\Gamma_{10x_3}^{Z} \mid \mid \Gamma_{00x_3}^{Z}  \right), \\
D_{Z}^{(2)}(x_3)& \triangleq & \mathbb{D} \left(\Gamma_{01x_3}^{Z} \mid \mid \Gamma_{00x_3}^{Z}  \right), \\
D_{Z-Y}^{(\ab{\ell})}(x_3) & \triangleq &  D_{Z}^{(\ab{\ell})}(x_3) - D_{Y}^{(\ab{\ell})}(x_3), \quad \forall \ab{\ell} \in \{1,2\}. \IEEEeqnarraynumspace
\end{IEEEeqnarray}
Also, we define for each $x_3 \in \mathcal{X}_3$ and $\rho_1,\rho_2\geq  0$: 
\begin{IEEEeqnarray}{rCl}
\chi^2(\rho_1,\rho_2, x_3) &\triangleq& \sum_{z \in \mathcal{Z}} \left[\frac{\rho_{1}}{\rho_{1} + \rho_{2}}\frac{ \Gamma_{10x_3}^{Z}(z)}{\Gamma_{00x_3}^{Z}(z)} \nonumber \right.\\ 
    &&\left.\hspace{.8cm}+ \frac{\rho_{2}}{\rho_{1}+ \rho_{2}} \frac{\Gamma_{01x_3}^{Z}(z)}{\Gamma_{00x_3}^{Z}(z) } -1\right]^2. \label{eq:def_xi_distance}\IEEEeqnarraynumspace
\end{IEEEeqnarray}


\medskip
\mw{\begin{theorem}
\label{main_theorem}
\highlight{Choose an arbitrary set of}
\begin{itemize}
    \item a positive integer $\tau$;
    \item a positive real number $\phi_1, \phi_2 \in [0,1]$;
    \item a joint distribution $P_{X_3T}$ over $\mathcal{X}_3 \times \{1,\ldots, \tau\}$;
    \item non-negative numbers $\{ \rho_{1,t}, \rho_{2,t}\}_{t=1}^\tau$;
    \item a non-negative sequence $\{\omega_n\}_{n=1}^\infty$ satisfying {\eqref{eq:lim_omega_n_def}}.
\end{itemize}
For any $\epsilon>0$, \ab{arbitrary small positive numbers  $\xi_m \in (0,1)$ for all $m \in \{1,\ldots,6\}$, and sufficiently large blocklength $n$, it is possible to find codes}, $\mathcal{C}_{1,t}, \mathcal{C}_{2,t}, \mathcal{C}_{3,t}$ of blocklengths \ab{$n_t=\lfloor n \cdot P_T(t)\rfloor$}, for $t=1, \ldots, \tau$, and message sizes 
\begin{IEEEeqnarray}{rCl}
\!\!\!\!\!\!\!\!\!\!\!\log(\mathsf{M}_1) &=& \phi_1  \cdot  (1-\xi_1)\omega_n \sqrt{n} \mathbb{E}_{P_{TX_3}} \left[ \rho_{1,T} D_{Y}^{(1)}(X_3) \right], \label{eq:th_1_log_m1} \\
\!\!\!\!\!\!\!\!\!\!\!\log(\mathsf{M}_2) &=&\phi_2 \cdot  (1-\xi_2)  \omega_n \sqrt{n} \mathbb{E}_{P_{TX_3}} \left[ \rho_{2,T} D_{Y}^{(2)}(X_3) \right], \label{eq:th_1_log_m2}\\
\!\!\!\!\!\!\!\!\!\!\!\log(\mathsf{M}_3) &=& (1-\xi_3) n I(X_3;Y \mid X_1=0, X_2=0, T).\label{eq:th_1_log_m3}\\
\!\!\!\!\!\!\!\!\!\!\!\log(\mathsf{K}_1) &=&\phi_1 \cdot  (1-\xi_4)  \omega_{n}\sqrt{n} \mathbb{E}_{P_{TX_3}}  \left[ \rho_{1,T} D_{Z-Y}^{(1)}(X_3) \right], \\ \label{eq:th_1_log_k1_k2}
\!\!\!\!\!\!\!\!\!\!\!\log(\mathsf{K}_2) &=& \phi_2 \cdot  (1-\xi_5) \omega_{n}\sqrt{n} \mathbb{E}_{P_{TX_3}}  \left[ \rho_{2,T} D_{Z-Y}^{(2)}(X_3) \right], \label{eq:th_1_log_k1_k2}
\end{IEEEeqnarray}
so that the encoding/decoding scheme described in the previous subsection achieves probability of error $P_e\leq \epsilon$ 
and 
average warden divergence
\begin{IEEEeqnarray}{rCl}
\lefteqn{
\frac{1}{\ab{\mathsf{M_3}}} \sum_{w_3=1}^{\ab{\mathsf{M_3}}} \delta_{n, w_3}} \nonumber \\
&\leq & \phi_1 (1+\highlight{\xi_6}) \frac{\omega_n^2}{2} \mathbb{E}\left[ (\rho_{1,T}+\rho_{2,T})^2 \cdot \chi^2(\rho_{1,T}, \rho_{2,T}, X_3)\right].\IEEEeqnarraynumspace
\end{IEEEeqnarray} 
\end{theorem}
\begin{IEEEproof} The  proof is omitted. It  follows from the schemes described in the previous sections~\ref{sec:coding_overview} and \ref{sec:gen}.
\end{IEEEproof}
}

The logarithmic scalings of the covert-message sizes $\mathsf{M}_1$ and $\mathsf{M}_2$  are at most square-root-$n$ scalings (because $\omega_n$ vanishes), indicating that the number of covert bits that can be transmitted is only in the order of square-root-$n$. However, we have the usual linear-in-$n$ \highlight{behavior} for the logarithmic scaling of the non-covert-message sizes. Communication of covert messages is thus of zero-rate while non-covert messages are communicated at standard  positive rates.  To obtain meaningful quantities, we will therefore define classical rates for User 3, while for the covert users we scale the logarithms of the message sizes by $\sqrt{n}$, and call the resulting asymptotic limits square-root rates. 

The \ab{secret-}key and covert-message square-root-scalings all depend on the vanishing sequence $\omega_n$. Increasing $\omega_n$ proportionally increases the permissible covert-message size but also quadratically increases the average divergence at the warden. To eliminate this dependence, we normalize the covert-message rates and the secret-key rates by the square-root of the average warden-divergence, leading to
\mw{
\begin{IEEEeqnarray}{rCl}
r_\ell & :=& \lim_{n\to \infty} \frac{\log(\mathsf M_\ell)}{\sqrt{{n \mathbb{E}_{W_3}\left[ \delta_{n, W_3}\right]}}}, \quad \ell \in\{1,2\}, \\ 
R_3 & :=& \lim_{n\to \infty} \frac{\log(\mathsf M_3)}{n}  \\ 
k_\ell &:=&  \lim_{n\to \infty} \frac{\log (\mathsf K_\ell)}{\sqrt{{n \mathbb{E}_{W_3}\left[ \delta_{n, W_3}\right]}}}, \quad \ell \in\{1,2\}.
\end{IEEEeqnarray}
}

We then obtain the following asymptotic capacity result for our setup with mixed covert and non-covert {users}. 
\begin{theorem}
\label{th:asymp_result}
There  exists a sequence of encodings and decodings \highlight{functions} satisfying 
\begin{subequations}\label{eq:Pd}
\begin{IEEEeqnarray}{rCl}
 \lim_{n \rightarrow \infty} \ab{P_{e,\mathcal{H}}}  &=& 0, \qquad \forall \ab{\mathcal{H}} \in \{0,1\}, \\
 \lim_{n \rightarrow \infty} \delta_{n,w_3} &=& 0, \qquad \forall w_3 \in \mathcal{M}_3,
 \end{IEEEeqnarray}
 \end{subequations}
 if, and only if, \mw{ for \ab{all} $\ell\in\{1,2\}$:
\begin{IEEEeqnarray}{rCl}
  r_{\ell}
  &=& \sqrt{2} \beta_{\ell} \frac{\mathbb{E}_{P_{TX_3}} \left[ \rho_{\ell,T} D_{Y}^{(\ell)}(X_3) \right]}{\sqrt{\mathbb{E}_{P_{TX_3}} \left[ \left( \rho_{1,T} + \rho_{2,T} \right)^2 \chi^2(\rho_{1,T}, \rho_{2,T}, X_3) \right]}}\label{eq:asymp1}, \nonumber \\ \\
 R_3 &\leq&  \mathbb{I}(X_3;Y \mid X_1=0,X_2=0,T)\label{eq:asymp3}, \\[1.2ex]
 k_{\ell}  &\geq& \sqrt{2}\beta_{\ell} \frac{\mathbb{E}_{P_{TX_3}}  \left[ \rho_{\ell,T} D_{Z-Y}^{(\ell)}(X_3) \right]} {\sqrt{\mathbb{E}_{P_{TX_3}} \left[ \left( \rho_{1,T} + \rho_{2,T} \right)^2 \chi^2({\rho}_{1,T}, \rho_{2,T}, X_3) \right]}}, \nonumber \\ \label{eq:asympkey} \IEEEeqnarraynumspace
\end{IEEEeqnarray}
for some pmf $P_{X_3T}$ over $\mathcal{X}_3 \times \{1,\ldots, \highlight{6}\}$, \highlight{positive} parameters $\{\rho_{1,t},\rho_{2,t}\}_{t=1}^\highlight{6}$, and $\ab{(\beta_1, \beta_2)}\in \ab{(0,1]}^2$.}
\end{theorem}
\begin{IEEEproof}
The ``if"-direction follows  from Theorem~\ref{main_theorem} \mw{ by choosing $\phi_1= \beta_1^2$  and $\phi_2=\beta_1 \beta_2$ \ab{when $\phi_1 \geq \phi_2$ and by setting $\phi_2= \beta_2^2$  and $\phi_1=\beta_1 \beta_2$ otherwise.}}
The ``only if"-direction is  sketched in Section~\ref{sec:coding}. 
\end{IEEEproof}
\subsection{Numerical examples}
\label{sec:simulations}
\begin{figure}[!h]
    \centering
    \begin{tikzpicture}
        \begin{axis}[%
            width=2.219in,
            height=0.802in,
            at={(0in,0in)},
            scale only axis,
            xmin=0,
            xmax=0.4,
            xlabel style={font=\color{white!15!black}},
            xlabel={Covert square-root-rate $r_2$},
            ymin=0,
            ymax=0.2,
            ytick={0, 0.1,  0.2},
            ylabel style={font=\color{white!15!black}},
            ylabel={Non-Covert rate $R_3$},
            axis background/.style={fill=white},
            title style={font=\bfseries},
            axis x line*=bottom,
            axis y line*=left,
            legend style={legend cell align=left, align=left, draw=white!15!black}
        ]
        \addplot [color=red]
            table[row sep=crcr]{%
                0.0209393603329447	0.197534050675693\\
                0	0.197534050675693\\
                0	0\\
                0.365874302371556	0\\
                0.365874302371556	0.00020475656488757\\
                0.341260993532471	0.0690075035111324\\
                0.33515527796459	0.0843432013787619\\
                0.323111967569048	0.114542756703675\\
                0.321752065485472	0.117541188433987\\
                0.315786299748673	0.129279386773892\\
                0.313386059962277	0.133311270079798\\
                0.304005523290595	0.149005425754013\\
                0.300323068543084	0.154598560290224\\
                0.283786335605196	0.175179901373978\\
                0.28137989410422	0.177415385091661\\
                0.272150823608734	0.184553522225264\\
                0.260392702464963	0.192644751530112\\
                0.255896812940455	0.193990325298559\\
                0.253087357283375	0.194657732732708\\
                0.251835506846633	0.194944103612356\\
                0.24821124899269	0.195608792433381\\
                0.241039965795286	0.196904759679634\\
                0.230998346134984	0.197390745020062\\
                0.178917199555822	0.197488598853102\\
                0.0209393603329447	0.197534050675693\\
                };
        
        \addplot [color=blue, dashed]
            table[row sep=crcr]{%
                0.18053796335818	0.19666811197991\\
                0	0.19666811197991\\
                0	0\\
                0.0159641442318079	0\\
                0.365564895078077	0\\
                0.366261291736838	0\\
                0.366261291736838	3.11518313179177e-09\\
                0.18053796335818	0.19666811197991\\
                };
        
        \end{axis}
        
    \end{tikzpicture}%
    
    \caption{Rate-region $(r_2,R_3)$ for secret-key rates $k_1\leq 0.8, k_2\leq 0.8$ and $r_1=0.5$ (solid line) and a degenerate region when restricting to  $|\mathcal{T}|=1$ (dashed line).}

    \label{fig:simulation_time_sharing_beneficial}
\end{figure}
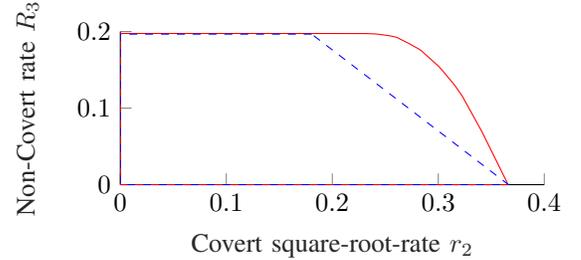

Consider binary input alphabets at all users, $\mathcal{X}_1=\mathcal{X}_2=\mathcal{X}_3=\{0,1\}$, and the following channels to the legitimate receiver and the warden. Here the various rows correspond to the different  triples  $(x_1,x_2,x_3)$ in lexicographic order and the  columns to the six $y$- or $z$-values.
\begin{IEEEeqnarray}{rCl}
    \highlight{\Gamma_{Y \mid X_1X_2X_3}}=\begin{bmatrix}
        \highlight{0.28} & 0.26 & 0.02 & \highlight{0.01} & 0.18 &  0.25 \\
        0.12 & 0.36 & 0.29 & 0.06 & 0.11 &  0.06 \\
        0.17 & 0.14 & 0.25 & 0.10 & 0.13 &  0.21 \\
        0.05 & 0.15 & 0.31 & 0.28 & 0.01 &  0.20 \\
        0.08 & 0.39 & 0.02 & 0.25 & 0.18 &  0.08 \\
        0.05 & 0.21 & 0.13 & 0.28 & 0.03 &  0.30 \\
        0.15 & 0.05 & 0.10 & 0.17 & 0.33 &  0.20 \\
        0.05 & 0.25 & 0.10 & 0.20 & 0.10 &  0.30 
    \end{bmatrix}, \nonumber \\  \label{eq:channels}
\end{IEEEeqnarray}
\begin{IEEEeqnarray}{rCl}
    \Gamma_{Z \mid X_1X_2X_3}= \begin{bmatrix}
        0.15 & 0.11 & 0.57 & 0.01 & 0.06 & 0.10 \\
        0.15 & 0.41 & 0.12 & 0.15 & 0.06 & 0.11 \\
        0.23 & 0.02 & 0.01 & 0.48 & 0.10 & 0.16 \\
        0.14 & 0.17 & 0.21 & 0.12 & 0.24 & 0.12 \\
        0.01 & 0.12 & 0.19 & 0.15 & 0.19 & 0.34 \\
        0.10 & 0.11 & 0.15 & 0.14 & 0.18 & 0.32 \\
        0.05 & 0.15 & 0.15 & 0.20 & 0.10 &  0.35 \\
        0.10 & 0.10 & 0.27 & 0.13 & 0.20 &  0.20
    \end{bmatrix}. \nonumber \\
\end{IEEEeqnarray}
Figure~\ref{fig:simulation_time_sharing_beneficial} illustrates the rate-region in Theorem~\ref{th:asymp_result} under the additional constraint on the secret-key \ab{rates $k_1\leq 0.8$, $k_2\leq 0.8$} (solid line) and the corresponding reduced rate-region when one imposes $T=1$ (dashed line), i.e., when  in our scheme
communication takes place only over a single phase. The obtained results prove that the covert capacity,  i.e., achievable square-root rates, is improved 
 when the users can communicate  over different phases and using different codes in the various phases.

\begin{figure}[!h]
    \centering
    \begin{tikzpicture}
        \begin{axis}[
            width=2.219in,
            height=0.902in,
            at={(0in,0in)},
            scale only axis,
            xmin=0,
            xmax=1,
            xlabel style={font=\color{white!15!black}},
            xlabel={Secret-key square-root-rate $k_2$},
            ymin=0,
            ymax=0.85,
            ylabel style={font=\color{white!15!black}},
            ylabel={Covert square-root-rate $r_2$},
            axis background/.style={fill=white},
            title style={font=\bfseries},
            axis x line*=bottom,
            axis y line*=left,
            legend style={legend cell align=left, align=left, draw=white!15!black}
        ]
       
            \addplot [color=black, dashdotted]
              table[row sep=crcr]{%
            0	0\\
            0.05	0.130115815930637\\
            0.1	0.260231631861274\\
            0.15	0.39034744779191\\
            0.2	0.520463263722547\\
            0.25	0.650579079653184\\
            0.3	0.780694895583821\\
            0.35	0.816679234764655\\
            0.4	0.816679234764655\\
            0.45	0.816679234764655\\
            0.5	0.816679234764655\\
            0.55	0.816679234764655\\
            0.6	0.818212418241385\\
            0.65	0.823688104809623\\
            0.7	0.823688104809623\\
            0.75	0.823688104809623\\
            0.8	0.823688104809623\\
            0.85	0.823688104809623\\
            0.9	0.823688104809623\\
            0.95	0.823688104809623\\
            1	0.823688104809623\\
            };
            
            \addplot [color=blue, dashed]
              table[row sep=crcr]{%
            0	0\\
            0.05	0.235812824459213\\
            0.1	0.235812824459213\\
            0.15	0.235812824459213\\
            0.2	0.235812824459213\\
            0.25	0.235812824459213\\
            0.3	0.235812824459213\\
            0.35	0.235812824459213\\
            0.4	0.235812824459213\\
            0.45	0.235812824459213\\
            0.5	0.235812824459213\\
            0.55	0.235812824459213\\
            0.6	0.235812824459213\\
            0.65	0.235812824459213\\
            0.7	0.235812824459213\\
            0.75	0.235812824459213\\
            0.8	0.235812824459213\\
            0.85	0.235812824459213\\
            0.9	0.235812824459213\\
            0.95	0.235812824459213\\
            1	0.235812824459213\\
            };
            
            \addplot [color=red, solid]
              table[row sep=crcr]{%
            0	0\\
            0.05	0.619621086839637\\
            0.1	0.658135475139432\\
            0.15	0.697745909281459\\
            0.2	0.7360782728842\\
            0.25	0.772984150029413\\
            0.3	0.812282179138629\\
            0.35	0.826375679242755\\
            0.4	0.82642859609202\\
            0.45	0.826461923639183\\
            0.5	0.827025876391132\\
            0.55	0.827025876391132\\
            0.6	0.82807250538062\\
            0.65	0.82807250538062\\
            0.7	0.82807250538062\\
            0.75	0.82807250538062\\
            0.8	0.82807250538062\\
            0.85	0.82807250538062\\
            0.9	0.82807250538062\\
            0.95	0.82807250538062\\
            1	0.82807250538062\\
            };
            
        \end{axis}
        
        \end{tikzpicture}%
        \caption{Covert rate $r_2$ as function of secret-key rate $k_2$ when optimizing over $P_{X_3T}$  (solid line) and when choosing $X_3=0$ or $X_3=1$ deterministically (dashed and dash-dotted lines) for a covert rate $r_1=0.1$ and a secret-key rate $k_1 \leq 0.8$.}
    
        \label{fig:non_covert_improves_covert_capacity}
\end{figure}
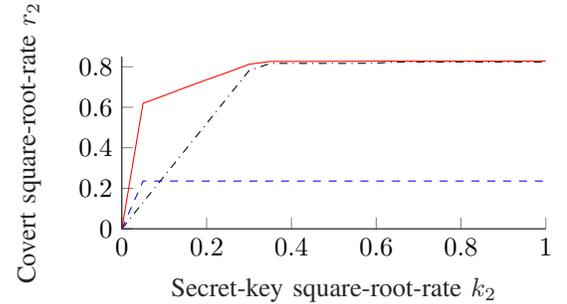

Figure~\ref{fig:non_covert_improves_covert_capacity} illustrates the maximum covert-user square-root rate $r_2$ as function of the secret-key rate \ab{$k_2$}, i.e., when one optimizes {$P_{X_3T}$} (solid line). This is compared to scenarios where non-covert User 3 sends constant symbols $X_3=0$ (dashed line) or $X_3=1$ (dash-dotted line). The observed performance underlines that the presence of non-covert User 3  can increase the covert capacity at Users 1 and 2.

\begin{figure}[!h]
    \centering
    \begin{tikzpicture}

        \begin{axis}[%
                width=2.219in,
                height=0.802in,
                at={(0in,0in)},
                scale only axis,
                xmin=0,
                xmax=0.7,
                xlabel style={font=\color{white!15!black}},
                xlabel={Covert square-root-rate $r_2$},
                ymin=0,
                ymax=0.2,
                ytick={0, 0.1,  0.2},
                ylabel style={font=\color{white!15!black}},
                ylabel={Non-Covert rate $R_3$},
                axis background/.style={fill=white},
                title style={font=\bfseries},
                axis x line*=bottom,
                axis y line*=left,
                legend style={legend cell align=left, align=left, draw=white!15!black}
            ]
            \addplot [color=black, dashed]
                  table[row sep=crcr]{%
                1.77972354293735e-07	0\\
                9.24722273239988e-07	0\\
                1.28262630379367e-06	0\\
                1.57923360902797e-06	0\\
                1.60757470770757e-06	0\\
                2.18672888627377e-06	0\\
                3.15583642498565e-06	0\\
                3.83636298066565e-06	0\\
                4.59420092057048e-06	0\\
                7.5713378953655e-06	0\\
                1.40857988992564e-05	0\\
                1.99248310622231e-05	0\\
                2.46620515194424e-05	0\\
                2.55525811790027e-05	0\\
                2.92903849235573e-05	0\\
                3.25239878985047e-05	0\\
                3.58609885302899e-05	0\\
                4.00805258409743e-05	0\\
                4.21970493268816e-05	0\\
                4.53365235073852e-05	0\\
                6.16242574539614e-05	0\\
                6.65426964019677e-05	0\\
                7.6132211504721e-05	0\\
                8.08610547761701e-05	0\\
                8.444004128534e-05	0\\
                8.49350001903799e-05	0\\
                0.000124951506738062	0\\
                0.000128391150556218	0\\
                0.000155430230610769	0\\
                0.000179657514605798	0\\
                0.000191938051779501	0\\
                0.000201248590660078	0\\
                0.000250742995779308	0\\
                0.000254541964520669	0\\
                0.000297255906973181	0\\
                0.000312520736969612	0\\
                0.000320553464335987	0\\
                0.00032362081232753	0\\
                0.000336000888571033	0\\
                0.000397675664332345	0\\
                0.000416549323891534	0\\
                0.000435737439046125	0\\
                0.000445675311085909	0\\
                0.000460841244903013	0\\
                0.000489764080076321	0\\
                0.000501782922922911	0\\
                0.000510895926054142	0\\
                0.00051858060671794	0\\
                0.000534712905862477	0\\
                0.000666379043312699	0\\
                0.000695182186310884	0\\
                0.000717654079503694	0\\
                0.000767938741204489	0\\
                0.000770215563780916	0\\
                0.000789075183164599	0\\
                0.000962149015119825	0\\
                0.00109105093870636	0\\
                0.00114189888547703	0\\
                0.00128112617726145	0\\
                0.0023200975624742	0\\
                0.00263042722092185	0\\
                0.00323838343781641	0\\
                0.00363097000859849	0\\
                0.00385667284300041	0\\
                0.00398281822127545	0\\
                0.00486018653987314	0\\
                0.00537741299713777	0\\
                0.00633287777156006	0\\
                0.00637146015259385	0\\
                0.00639421766952885	0\\
                0.00641071654294489	0\\
                0.0072612319237656	0\\
                0.00802805316291177	0\\
                0.00851102878364846	0\\
                0.0198118015321473	0\\
                0.0342375279487474	0\\
                0.0475080531116297	0\\
                0.0538610694060557	0\\
                0.0612948314141095	0\\
                0.0727582091253656	0\\
                0.0786112146758962	0\\
                0.0786112146758962	0.000350376841645943\\
                0.0565228878692079	0.0642091201256181\\
                0.0535555881184512	0.0723088020210261\\
                0.0516658364032987	0.0771219689519001\\
                0.0472256598296449	0.0847885317389099\\
                0.0409356055296297	0.0902790668071149\\
                0.0121423776174048	0.101923909435917\\
                0.000270120219338757	0.106176748012991\\
                0	0.106176748012991\\
                0	0\\
                3.05863898172023e-10	0\\
                3.23943431192293e-10	0\\
                5.36356586068584e-10	0\\
                7.31143713721128e-10	0\\
                9.05064447941155e-10	0\\
                7.18048434527286e-09	0\\
                2.36782437434929e-08	0\\
                1.66047565735782e-07	0\\
                1.77972354293735e-07	0\\
                };
                
                \addplot [color=blue, dashdotted]
                  table[row sep=crcr]{%
                0.0209269421595458	0.197534055988625\\
                0	0.197534055988625\\
                0	0\\
                0.366260167743622	0\\
                0.366260167743622	5.92671339086962e-07\\
                0.352467830673336	0.0424563176191503\\
                0.343939259498895	0.0660921094158863\\
                0.339181407707355	0.0787362566994103\\
                0.331392190828389	0.0975435064355721\\
                0.328772634590651	0.103252582194686\\
                0.323773632757054	0.113946412617713\\
                0.321649663783969	0.118235190707539\\
                0.319782634959482	0.121845752369862\\
                0.30705598126345	0.14526089851093\\
                0.302077024896372	0.152408280744081\\
                0.300104990775399	0.15507733345502\\
                0.295096021914815	0.161840713033189\\
                0.289490747688808	0.168581308563562\\
                0.286194502639452	0.17252017700587\\
                0.283322580590531	0.175899359820123\\
                0.277828306137395	0.180667197252443\\
                0.276936739926482	0.181400071218315\\
                0.271389569503115	0.185843088156316\\
                0.268062365695835	0.188216602249852\\
                0.26650468880651	0.18917940579763\\
                0.259215929414188	0.193180954794483\\
                0.25905367335321	0.193251437416159\\
                0.256833409554561	0.194128357322257\\
                0.249730705704971	0.195836584992729\\
                0.241703581766645	0.196886110493879\\
                0.230366610495996	0.197459034796656\\
                0.225782818358957	0.197506105910677\\
                0.203832002692384	0.197526401063909\\
                0.0209269421595458	0.197534055988625\\
                };
                
                \addplot [color=red, solid]
                  table[row sep=crcr]{%
                0.479011611731705	0.197533997476211\\
                0.410539988152724	0.19753483958795\\
                0	0.19753483958795\\
                0	0\\
                0.653523154298476	0\\
                0.653523154298476	3.12351191284259e-07\\
                0.632458823538284	0.0595218125644257\\
                0.616857973913381	0.1012818747965\\
                0.614658440332804	0.107035517437211\\
                0.602722944584103	0.130441891300352\\
                0.594119614686427	0.144833010129501\\
                0.588539286489944	0.153590532618445\\
                0.579399623769747	0.165505157364048\\
                0.575144712444228	0.170503386973003\\
                0.573578728721614	0.172269156043192\\
                0.566786043546573	0.179332445370669\\
                0.55924302652555	0.18516940617372\\
                0.5525498908793	0.189831822875043\\
                0.547480271942865	0.192594783879944\\
                0.539083280295056	0.195634102869646\\
                0.536143631847814	0.196386133343222\\
                0.53495628444032	0.196651240074962\\
                0.531392617043895	0.197106424759261\\
                0.529525250269319	0.197309363015386\\
                0.528705907572764	0.197390787520215\\
                0.527300550373405	0.197463014950631\\
                0.525123037001012	0.197502586668147\\
                0.522032274031789	0.197513593323215\\
                0.516431893237586	0.197527890018186\\
                0.508952016438069	0.197533488075582\\
                0.479011611731705	0.197533997476211\\
                };

        \end{axis}
    \end{tikzpicture}%
    \caption{Rate-region $(r_2,R_3)$ \ab{for secret-key rates $k_1\leq 0.8, k_2\leq 0.8$} and different rates: \ab{$r_1=0.75$} (dashed line), $r_1=0.5$ (dash-dotted line) and \ab{$r_1=0.25$} (solid line).}
    \label{fig:simulation_r2_vs_r3_for_different_rate_r1}
\end{figure}
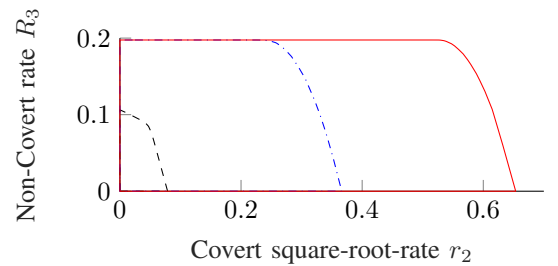

Figure~\ref{fig:simulation_r2_vs_r3_for_different_rate_r1} illustrates the rate-region  in Theorem~\ref{th:asymp_result} at different rates $r_1$ for the covert User 1, showcasing the trade-off between the different users \ab{at secret-key rates $k_1\leq 0.8$ and $k_2\leq 0.8$}.
\vspace{-5mm}
\vspace{-8mm}
\vspace{-8mm}
\vspace{-2mm}
\section{Proof of Converse to Theorem~\ref{th:asymp_result}}\label{sec:coding}
\mw{Fix a sequence of encodings and decodings satisfying \eqref{eq:Pd}. 
Recalling the definition of $  \widehat{Q}_{\mathcal{C}, w_3}^{n}(z^{n})$ in \eqref{eq:def_Q_C_w3} and denoting the $i$-th component of \ab{the} codeword $x_3^n(w_3)$ by $\ab{x_{3,i}(w_3)}$, we obtain on average over the random code and  message $W_3$: 
\begin{align}
    &\mathbb{E}[\delta_{n,W_3}] \nonumber \\
    &\overset{(a)}{=} \sum_{i=1}^n \mathbb{E}_{W_3} \left[ \mathbb{D} \left( \widehat{Q}_{\mathcal{C},w_3}^{(i)} \| \highlight{\Gamma_{Z \mid X_1X_2X_3}}(\cdot| 0, 0, \ab{x_{3,i}}(W_3))] \right)\right] \label{eq:a} \\
    &\overset{(b)}{\geq} n \mathbb{E}_{P_{TX_3}} \Bigg[ \frac{\left(\alpha_{n,T,1} + \alpha_{n,T,2} \right )^2}{2} \chi^2(\ab{\alpha_{n,T,1}}, \ab{\alpha_{n,T,2}},\ab{X_{3,T}})  \nonumber \\
    & \hspace{30mm}+ o\left(\max_\ab{\ell \in \{1,2\}} \{ \alpha_{n,T,\ab{\ell}}  \} \right) \Bigg], \label{eq:converse_upper_bound_divergence_final}
\end{align}
where  $T$ is uniform over $\{1,\ldots,n\}$. Here, (a) holds by the memoryless nature of the channel and by defining $\widehat{Q}_{\mathcal{C},w_3}^{(i)}$ as the $i$-th marginal of $\widehat{Q}_{\mathcal{C},w_3}^{n}$; (b) holds by an extension of~\cite[Lemma 1]{bloch_k_users_mac}, and upon defining  
\begin{equation}
    \label{eq:def_alpha_t_converse}
    \alpha_{n,i,\ell} \triangleq \frac{1}{\mathsf{M}_\ell \mathsf{K}_\ell} \sum_{w_\ell=1}^{\mathsf{M}_\ell} \sum_{s_\ell=1}^{\mathsf{K}_\ell} \mathds{1} \{ x_{\ell,i}(w_\ell,s_\ell ) = 1\}, \quad   \ell \in \{1,2\}.
\end{equation}
Notice that by \eqref{eq:a} each $\alpha_{n,i, \ell} \to 0$ as $n\to \infty$. 
And thus by standard arguments and an  extension of~\cite[Lemma 2]{bloch_k_users_mac}:
\begin{IEEEeqnarray}{rCl}
   \log(\mathsf{M}_1) & \leq &n \mathbb{E}_{P_{TX_3}} \left[ \alpha_{n,T,1} \ab{D_{Y}^{(1)}}(\ab{X_{3,T}}) + o(1) \right]  + 1,  \label{eq:converse_bound_m1} \IEEEeqnarraynumspace
\end{IEEEeqnarray}
and similarly 
\begin{IEEEeqnarray}{rCl}\label{2}
\!\!\!\!\!\! \log(\mathsf{M}_2)  &\leq & \highlight{n} \mathbb{E}_{P_{TX_3}} \left[ \alpha_{n,T,2} \ab{D_{Y}^{(2)}}(\ab{X_{3,T}}) + o(1) \right]  + 1.
\end{IEEEeqnarray}
Moreover for all $\ell \in\{1,2\}$,
\begin{IEEEeqnarray}{rCl}
    \log(\mathsf{M}_{\ell} \mathsf{K}_{\ell}) &\geq& n \mathbb{E}_{P_{TX_3}} \left[\alpha_{n,T,\ell} \ab{D_{Z}^{(\ell)}}(\ab{X_{3,T}}) +o(1) \right] \label{eq:converse_bound_mlkl}\IEEEeqnarraynumspace
\vspace{-0.1cm}
\end{IEEEeqnarray}
Define next 
\begin{equation}
\rho_{n,T,\ell} \triangleq \frac{\alpha_{n ,T,\ell}}{ \mathbb{E}[ \alpha_{n,T,1}+\alpha_{n,T,1} ] }, \quad \ell\in\{1,2\},
\end{equation} and notice that by 
\eqref{eq:converse_upper_bound_divergence_final}, \eqref{eq:converse_bound_m1}, and \eqref{2}:
\begin{IEEEeqnarray}{rCl}
\lefteqn{  \frac{  \log(\mathsf{M}_\ell) }{ \sqrt{ n \mathbb{E}_{W_3}[\delta_{n,W_3}] }}}\nonumber\\
& =&   \frac{ \beta_\ell \mathbb{E}_{P_{TX_3}} \left[ \rho_{n,T,\ell} \ab{D_{Y}^{(\ell)}}(\ab{X_{3,T}}) \right]}{ \mathbb{E}_{P_{TX_3}} \Bigg[ \frac{\left(\rho_{n,T,1} + \rho_{n,T,2} \right)^2}{2} \chi^2(\ab{\rho_{n,T,1}}, \ab{\rho_{n,T,2}},\ab{X_{3,T}}) \Bigg]}+o(1), \nonumber\\
\vspace{-0.2cm}
\end{IEEEeqnarray}
\vspace{-0.1cm}
for some $\beta_\ell\in[0,1]$. Combined with  \eqref{eq:converse_bound_m1}--\eqref{2}, this yields:
\begin{IEEEeqnarray}{rCl}
\lefteqn{\sqrt{ n\mathbb{E}_{W_3}[\delta_{n,W_3}]} \leq } \nonumber \\
&& \frac{n}{ \beta_\ell} \mathbb{E}_{P_{TX_3}} \Bigg[ \frac{\left(\rho_{n,T,1} + \rho_{n,T,2} \right)^2}{2} \chi^2(\ab{\rho_{n,T,1}}, \ab{\rho_{n,T,2}},\ab{X_{3,T}}) \Bigg]  \nonumber \\ && \hspace{0.7cm} +o(1)
\end{IEEEeqnarray}
which can be combined with \eqref{eq:converse_bound_mlkl}  to establish that 
\begin{IEEEeqnarray}{rCl}
\lefteqn{  \frac{  \log(\mathsf{M}_\ell  \mathsf{K}_{\ell}) }{ \sqrt{ n \mathbb{E}_{W_3}[\delta_{n,W_3}] }}  \geq } \nonumber \\
&& \beta_\ell  \frac{ \mathbb{E}_{P_{TX_3}} \left[ \rho_{n,T,\ell} \ab{D_{Z}^{(\ell)}}(\ab{X_{3,T}}) \right]}{ \mathbb{E}_{P_{TX_3}} \Bigg[ \frac{\left(\rho_{n,T,1} + \rho_{n,T,2} \right)^2}{2} \chi^2(\ab{\rho_{n,T,1}}, \ab{\rho_{n,T,2}},\ab{X_{3,T}}) \Bigg]}+o(1). \nonumber \\
\vspace{-0.3cm}
\end{IEEEeqnarray}
By standard arguments and since the probabilities that $X_{1,T}$ and $X_{2,T}$ differ from 0 vanish as $n\to \infty$:
\vspace{-0.15cm}
\begin{equation}
\frac{1}{n}    \log(\mathsf{ M}_3)  \leq I(X_{3,T}; Y_T| \ab{X_{1,T}=0,} X_{2,T}=0 \ab{, T})+o(1).
\end{equation}
Combining all arguments establishes the  converse result.}

\section{Summary and Discussion}
\label{sec:conclusion}
This paper establishes the fundamental limits of a multi-access communication setup with two covert users and one non-covert user communicating to the same receiver in presence of a warden. 
Both covert users also share a common secret-key of fixed key rate with the receiver. 
Our results highlight that multiplexing different codes in different phases is crucial to exhaust the entire tradeoff of achievable covert and non-covert rates.
Moreover, our results also show that the presence of the non-covert user can potentially improve the covert-capacity under a stringent \ab{secret-}key rate constraint.

\highlight{In a straightforward way, our results can also be extended to} multiple users with arbitrary finite alphabets (i.e., $\mathcal{X}_1$ and $\mathcal{X}_2$ not necessarily  binary).
\bibliographystyle{IEEEtran}
\bibliography{references}
\clearpage

\end{document}